\documentclass[a4paper, amsfonts, amssymb, amsmath, reprint, showkeys, nofootinbib, twoside]{revtex4-1}
\usepackage[english]{babel}
\usepackage[utf8]{inputenc}
\usepackage[colorinlistoftodos, color=green!40, prependcaption]{todonotes}
\usepackage[pdftex, pdftitle={Article}, pdfauthor={Author}]{hyperref} % For hyperlinks in the PDF
\usepackage{amstext,amssymb,amsfonts,amsmath,amsthm,mathtools}
\usepackage{graphicx,subfigure}
\usepackage{array}
\usepackage{epstopdf}
\usepackage{ulem}
\usepackage[font=small]{caption}
\usepackage{xcolor}
\usepackage{soul}
\usepackage{enumerate}
\usepackage[toc,page]{appendix}
\usepackage[margin=1.2in]{geometry}
\usepackage{environ}
\usepackage{url}
\usepackage{hyperref}

\NewEnviron{SMALL}{\scalebox{0.90}{$\BODY$}}
\graphicspath{ {./figures/} } % to tell LaTeX the images location folder
\begin{document}
\title{Chern-Simons approach to atmospheric dynamics}

\author{Mart\'{\i}n Jacques-Coper}
    \email[Email address: ]{mjacques@dgeo.udec.cl}% Your name
    \affiliation{Departamento de Geof\'{\i}sica \& Center for Climate and Resilience Research, Universidad de Concepci{\'o}n, Casilla 160-C, Concepci{\'o}n, Chile}
\author{Valentina Ortiz}
    \email[Email address: ]{vortiz@cecs.cl}% Your name
    \affiliation{Centro de Estudios Cient\'{\i}ficos (CECs), Av. Arturo Prat 514, Valdivia, Chile}
\author{Jorge Zanelli}
    \email[Email address: ]{z@cecs.cl}% Your name
    \affiliation{Centro de Estudios Cient\'{\i}ficos (CECs), Av. Arturo Prat 514, Valdivia, Chile}
%\date{\today} % Leave empty to omit a date

\begin{abstract}
A simplified model for a planet's atmosphere as an open two-dimensional Chern-Simons system is presented. The dynamical variables describe an ideal gas with velocity $\vec{v}$, density $\rho$, temperature $T$ and pressure $P$. Radiation exchange, diffusion and mechanical dissipation are also included. The resulting dynamics is given by a set of nonlinear differential equations of first order in time for the densities of energy, momentum and mass. The initial value problem can be integrated given the radiation balance of the planet. If the nonlinearities could be neglected the integration can be done in analytic form and small nonlinearities can be incorporated as perturbative corrections.
\end{abstract}
\maketitle

%%%%%%%%%%%%%%%%%%%%%%%%%%%%%%%
%\section{Introduction}  %  1  %
%%%%%%%%%%%%%%%%%%%%%%%%%%%%%%%
The importance of the long term behavior of the Earth's atmosphere cannot be overemphasized. In order to address this complex problem, having a robust working model that describes the time evolution of the atmosphere at the global scale would be ideal. Roughly 75\% of the atmosphere's mass is contained in the Troposphere, a layer $\sim$10 kilometers thick extending over a surface four-thousand times larger which, to a good approximation can be modelled as a two-sphere. At large scale, then, the Earth's atmosphere is essentially two-dimensional. Horizontal motions are ~3 orders of magnitude greater than vertical motions, which, nonetheless are of critical importance to atmospheric dynamics \cite{Wallace}. As an idealized exercise, in this note we consider a two-dimensional inviscid fluid on a rotating, perfectly spherical surface, subject to the inflow of energy coming from an external source representing the Sun. We describe the system by Chern-Simons (CS) equations in two spatial and one time dimensions \cite{Deser-Jackiw}. In their simplest form, the CS equations are first order linear differential equations for a three-component field theory. CS models have been proposed to describe other 2+1 dimensional systems like, for instance, in the fractional quantum Hall effect, anyons, 2+1 vortices \cite{Dunne}, or the electron gas in the two-dimensional planes of high temperature superconductors \cite{RandjbarDaemi,Sedrakyan,Wang}, gravity in a 2+1 dimensional spacetime \cite{Witten} or graphene \cite{AVZ}.

In the two-dimensional approximation, effects such as the vertical features of the fluid are ignored. This could be viewed as a simplified approach than that of Charney's model, which describes the quasi-two-dimensional nature of the atmosphere as a quasi-geostrophic system \cite{Charney1971}. The large horizontal-to-vertical scale ratio has also been used to study the large-scale behavior of the atmosphere  as a two-dimensional homogeneous isotropic turbulent flow \cite{Boer-Shepherd1983}. Several experiments have also used this approximation (see, e.g. \cite{Afanasyev-Wells2005}). 

We assume the atmospheric fluid described by a velocity field $\vec{v}(t,\vec{x})$, mass density $\rho(t,\vec{x})$, pressure $P(t,\vec{x})$ and temperature $T(t,\vec{x})$, where $t$ is time and $\vec{x}$ indicates a position on the two-sphere. In an idealized situation, the dynamics of the system could be described as a compressible ideal gas under the influence of an energy source describing the Sun, and the Coriolis force due to the Earth's rotation. In addition, we assume the two-dimensional fluid to have constant specific heat and compressibility. We will also allow for slight dissipation and diffusion, so that in the absence of external influences, the system would relax towards a static uniform equilibrium configuration with $\vec{v}=0$, and constant $\rho$, $P$ and $T$.

One can expect that this idealized model, where nonlinearities are initially neglected, to be a reasonable approximation to describe the layers of mid- to high- troposphere. A better approximation could be obtained at a later stage by considering nonlinear effects as small perturbative corrections or in a fully nonlinear regime for larger ones. Also, a more accurate description could be expected in a multi-layer model, but for concreteness, here we concentrate in the simpler case of a single layer.

%%%%%%%%%%%%%%%%%%%%%%%%%%%%%%%%%%%%%%%%%%%%%%%%%%%
%\section{Chern-Simons single layer model}  % 2  %
%%%%%%%%%%%%%%%%%%%%%%%%%%%%%%%%%%%%%%%%%%%%%%%%%%%
Consider the flow of a single atmospheric layer described by a three-component vector $A_\mu=(A_0, A_1, A_2)$ that encodes information about the velocity, density and pressure of a two-dimensional fluid in a three-dimensional spacetime $\mathcal{M}=\mathbb{R}\times S^2$. The CS dynamics for $A_\mu$ takes the form of the system of equations
\begin{align} \label{CSeqs-0}
\partial_\mu A_\nu - \partial_\nu A_\mu = \epsilon_{\mu \nu \lambda} j^\lambda\,.
\end{align}
Here $j^\mu=(j^0, j^1, j^2)$ represents the effect of the interaction with external sources of energy and matter. These interactions include solar radiation --with its variations due to Earth rotation and orbital motion--, the heat emitted to outer space and also the influx of matter from the planet's surface and losses through precipitation or convection. The incoming and outgoing energy and matter fluxes must be such that the long-term integrated fluxes almost exactly cancel out. 

If we postulate the identification $A_i = \rho v_i=p_i$ (momentum density), and $A_0 = -P$ (negative pressure), the CS equations become
\begin{align} \label{CSeqs-1}
\partial_t(p_i) +\partial_i P & = \epsilon_{ik}j^k\, ,\\ \label{CSeqs-2}
\hat{r}\cdot [\nabla \times (\vec{p})] &= j^0\, .
\end{align}
The first equation is identified as Newton's second law in an instantaneous locally co-moving frame, $d\vec{p}/dt= \vec{F}$. In a rotating frame, the driving force on the right hand side of \eqref{CSeqs-1} can be identified with the Coriolis effect, $2\vec{\omega}\times \vec{p}$. We neglect the direct friction exerted on the fluid by the planet's surface, an approximation that may be correct for the higher layers of the Earth's atmosphere but not for the lower strata. 

The second equation relates the curl of $\vec{p}$ to some external current $j^0$, which represents the rate of change of matter density and therefore can be interpreted as the net rate of change of $\rho$. Hence, \eqref{CSeqs-2} establishes a connection between the curl of the momentum density $\vec{p}$ and $d\rho/dt$. This suggests interpreting $\nabla \times \vec{p}$ as a source for the continuity equation,
\begin{equation} \label{continuity}
\partial_t \rho + \nabla \cdot \vec{p} -\sigma \nabla^2 \rho = \beta \hat{r}\cdot [\nabla \times \vec{p}] \, ,
\end{equation}
where we have also included a diffusive term to allow the system to relax into a homogeneous distribution of matter in absence of a driver.

Clearly, the CS equations are not sufficient to describe the atmosphere, in particular, because it is an open system that receives and emits energy in the form of radiation. The rate of change of the internal energy is proportional to the energy entering the system.

%%%%%%%%%%%%%%%%%%%%%%%%%%%%%%%%%%%%%%%%%%%
%\subsection{Thermal energy balance} % 2.1 %
%%%%%%%%%%%%%%%%%%%%%%%%%%%%%%%%%%%%%%%%%%%
The ideal gas assumption implies that its temperature and pressure could be related through an equation of state of the form
\begin{equation}\label{eq:stateeq} 
P(t,x)= \alpha \rho(t,x) T(t,x) \equiv \alpha \tau(t,x) \,, 
\end{equation}
where $\alpha$ is the ideal gas constant. Substituting this equation in \eqref{CSeqs-1} relates the change in momentum of the fluid to the gradient of the thermal density $\tau\equiv \rho T$. 

The net amount of heat deposited on the atmosphere per unit area per unit time, $E(t,x)$, includes radiation directly from the Sun as well as that reflected by the surface and that emitted to outer space. In addition, heat also changes by diffusion as it flows from warmer to colder regions. Hence, we postulate that in a locally co-moving frame, the thermal energy density changes as
\begin{equation}\label{eq:heat-transfer}
\partial_t \tau - k\nabla^2 \tau = \frac{1}{c}E(t,x)\,,
\end{equation}
where $k$ is the diffusion coefficient and $c$ is the specific heat of the fluid that we assume to be constant.

Equations \eqref{CSeqs-1} and \eqref{eq:heat-transfer} are in an instantaneous co-moving frame. In order to translate the results to a reference frame fixed on the Earth, one should replace $\partial_t$ by the material derivative $\partial_t + (\vec{v}\cdot \nabla)$, which takes into account the local changes in the parameters of the fluid due to drifting. Putting all this information together, we arrive at the equations 
\begin{align}\label{tau} 
&\left(\partial_t - k \nabla^2 + \vec{v}\cdot \nabla \right) \tau = \frac{1}{c}E(t,\vec{x}) \;, \\
\label{p}
&\left(\partial_t +\eta \; +\; 2\vec{\omega} \times\; +\; \vec{v}\cdot \nabla \;\right) \vec{p} =  -\alpha \nabla \tau\;, \\
\label{ro}
&\left(\partial_t - \sigma \nabla^2 \right)\rho = \beta\, \hat{r}\cdot( \nabla \times \vec{p}) - \nabla \cdot \vec{p} \;, 
\end{align}
where all vectors except $\hat{r}$ are tangent to the sphere, and we have added a dissipative term to the momentum density equation, with damping coefficient $\eta$.

These coupled differential equations may be compared with more standard forms where the vertical dimension is not ignored, as for example in \cite{Lorenz,Vallis}, (see \cite{JC-O-Z}). This set can be numerically solved given $E(t,\vec{x})$ and the initial data $\tau(0,\vec{x})$, $\vec{p}(0,\vec{x})$, $\rho(0,\vec{x})$. The problem is greatly simplified if the advective term $\vec{v}\cdot \nabla$ is dropped. Removing it from \eqref{tau} allows solving it for $\tau$, which allows solving \eqref{p} for $\vec{p}$, which in turn determines $\rho$; dropping $\vec{v}\cdot \nabla$ from \eqref{tau} and \eqref{p}, reduces the system to a linear problem, which allows a complete analytic solution by Green's methods if $E(t,\vec{x})$ is known. In what follows we discuss and simulate this latter simplified model.

%%%%%%%%%%%%%%%%%%%%%%%%%%%%%%%%%%%%%%%%%%%%%%%%%%%%%%%%%%%
%\section{General solution of the linear problem} % 3 %
%%%%%%%%%%%%%%%%%%%%%%%%%%%%%%%%%%%%%%%%%%%%%%%%%%%%%%%%%%%

Consider a single-layer fluid governed by Eqs. (\ref{tau}-\ref{ro}), neglecting the advective terms, which could be added perturbatively later on. The solution has the form of a particular inhomogeneous part, plus a homogeneous term fitted to match the initial conditions or the equilibrium state in absence of drivers. We will take an idealized situation where the surface features (oceans, continents, ice covered regions, mountains, clouds, etc.) are ignored; the energy provided by the Sun is constant and uniformly distributed. We assume $E$ to be the difference between the absorbed ($E_{in}$) and emitted ($E_{out}$) contributions as
\begin{equation} \label{Ein}
E_{in}(t,\theta,\phi) = \mathcal{E}_{0}\,\hat{r} \cdot \hat{s} \; \Theta(\hat{r} \cdot \hat{s})\,, 
\end{equation}
where $-\pi/2\leq \theta \leq \pi/2$, $0\leq \phi \leq 2\pi$,  $\hat{r}\cdot \hat{s} = \cos{\Delta} \cos{\theta} \sin{(\phi +\omega t)} + \sin{\Delta} \sin{\theta}$, $\mathcal{E}_{0}$ is some fraction of the solar radiative flux density, $\Theta$ is the Heaviside step function, $\hat{r}$ the unit vector in the radial direction, $\hat{s}$ is the unit vector in the direction of the Sun, $\Delta$ is the tilt angle between the spin axis and the normal to the ecliptic plane and $\omega$ is the rotation (spin) frequency. The emitted energy is taken to be in a form similar to black-body radiation, $E_{out} = \lambda \tau(t,\theta,\phi)$, where $\lambda$ is an emissivity coefficient \cite{F1}. With this information, the thermal density can be found as
\begin{equation} \label{tau-Green}
\begin{split}
\tau(t,\vec{x}) = & \int d\vec{x}'\int_0^t  dt' G_k(t,t';\vec{x},\vec{x}')e^{-\frac{\lambda}{c}(t-t')}\, \\
& \times\frac{1}{c} E_{in}(t',\vec{x}')+ \tau(0,\vec{x})e^{-\frac{\lambda}{c}t} \;,
\end{split}
\end{equation}
where the Green function is given by
\begin{align} \label{eq:Gd}
\begin{split}
G_{k}(t,t';&\vec{x},\vec{x}') = \sum_{l=0}^{\infty}\sum_{m=-l}^{l}g_{k,l}(t,t')\\
    &\times Y_{l m}^*(\theta',\phi')\,Y_{l m}(\theta,\phi)\,,
\end{split}\\
g_{k,l}(t,t') &= \,\Theta(t-t')\,e^{-k l (l+1)(t-t')}\,,
\end{align}
where $Y_{l m}(\theta,\phi)$ are the spherical harmonics. Equating the integral of \eqref{Ein} to the emitted energy in a period of rotation, gives the global average equilibrium temperature $T_0 $ 
\begin{equation} \label{T0}
T_0 = \frac{\mathcal{E}_0}{4\lambda\rho_0}\;.
\end{equation}

\begin{figure}[h!]
\centering
\subfigure[]{
\includegraphics[width=7.2cm]{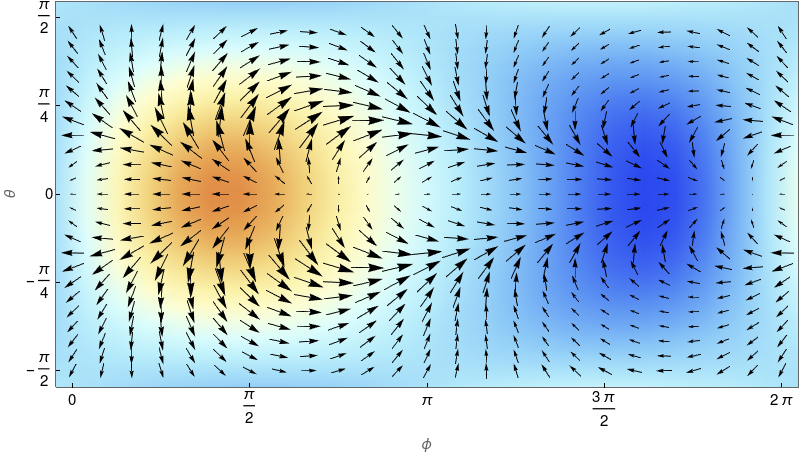}}
\subfigure[]{
\includegraphics[width=7.2cm]{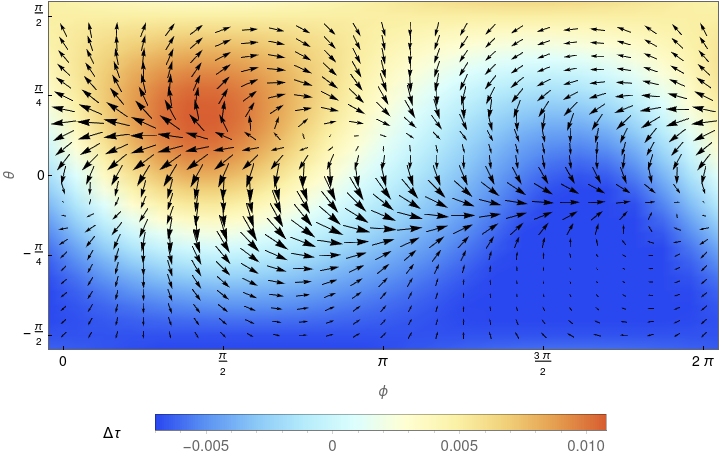}}
\caption{(a) Momentum density field $\vec{p}$ with background colors showing the thermal density distribution $\tau$ for (a) $\Delta=0^o$ and (b) $\Delta=23^o$. These plots use parameters $\mathcal{E}_0=0.02$, $c=1.1$, $k=0.3$, $\alpha=0.3$, $\eta=0.5$, $\sigma=0.3$, $\beta=0.9$ and $\lambda=0.005$. }		
\label{fig:fields}	
% It can be observed that $\rho$ varies by about 1\%, which would justify taking it as a constant.
\end{figure}

The analytic integration for $\vec{p}$ uses $\tau$ as input and the Green function for a damped harmonic oscillator,
\begin{equation} \label{p_i}
\begin{split}
\vec{p}(t,\vec{x}) = -\alpha \int_{-\infty}^{\infty}&\frac{e^{-\eta\,(t-t')}\sin[\nu\,(t-t')]}{\nu}\\
    & \times \nabla \tau(t',\vec{x}) \,dt'\,,
\end{split}
\end{equation} 
where $\nu=2\omega \sin{\theta}$ is the Foucault precession frequency. Finally, using this solution for $\vec{p}$, the matter density is obtained as
\begin{equation} \label{Int-rho}
\begin{split}
\rho(t,\vec{x}) =  \int d\vec{x}'& \int_0^t  dt' G_{\sigma}(t,t';\vec{x},\vec{x}') \\
    & \begin{aligned}
    \times F(t',\vec{x}') + \rho_0\,,
    \end{aligned}
\end{split}
\end{equation}
where $G_{\sigma}$ has the same form of \eqref{eq:Gd} replacing $k$ for $\sigma$, $\rho_0$ is the equilibrium value for the density and the source function $F$ in convolution with the Green function is
\begin{equation} \label{source_rho}
\begin{split}
F(t,\theta,\phi)= \beta &\frac{\partial_{\theta}(\cos{\theta}p^{\phi})-\partial_{\phi}p^{\theta}}{R\cos{\theta}} \\
& - \frac{\partial_{\theta}(\cos{\theta}p^{\theta}) + \partial_{\phi}p^{\phi}}{R\cos{\theta}}\,.
\end{split}
\end{equation}
The details of the integration outlined above can be found in \cite{JC-O-Z} and are included as supplemental material. The plots of $\tau$, $\vec{p}$ and $\rho$ for some representative values of parameters are plotted in Figure \ref{fig:fields}. The effect of varying those parameters can be explored using the Python interactive code in \cite{App}.

% % % % % FIGURE % % % % % 
% % % % % % % % % % % 
{\bf Comments:}
{\bf 1.} Equations (\ref{tau}-\ref{ro}) resemble those obtained by E. N. Lorenz in 1948 \cite{Lorenz} or, in the current approach to atmosphere dynamics, as for example in \cite{Vallis}. A natural question then is under what assumptions our model can be related to the standard equations. We address this question in \cite{JC-O-Z}, where we show that the corresponding equations match if the effective parameters $\eta, k, \sigma$ are properly identifed. \\
\noindent
{\bf 2.} The CS assumption suggests taking the thermal density ($\tau$), matter density ($\rho$ ), and momentum density ($\vec{p}$) as independent variables. The linearized equations decouple and are easily integrable. Identifying $\nabla \times \vec{p}$ as proportional to the flux of matter into the system is a hypothesis that could be tested by analyzing the effect of changing the value of $\beta$ in the solution, which affects the distribution of $\rho$.\\
\noindent
{\bf 3.} For the set of parameters chosen in the simulations --which might be pertinent for the Earth--, $\rho$ turns out to be quite uniform, varying by less than 1\% of its  entire range. In this case, the fields $\vec{v}$ and $\vec{p}$ differ roughly by a constant factor, and the same would be true of $T$ and $\tau$.\\
\noindent
{\bf 4.} The energy function $E(t,\vec{x})$ is the main input of the model. The parameters $\mathcal{E}_0, c, k, \alpha, \eta, \sigma, \beta, \lambda, \Delta$ can be adjusted to describe different features of our atmosphere or that of a different planet. In our simulation, we chose an $E(t,\vec{x})$ that corresponds to a fraction of the energy from the sun, ignoring the local differences in albedo or other geographic effects.\\
\noindent
{\bf 5.} The simulated fields resemble the migrating components of thermally driven atmospheric diurnal tides that occur due to solar heating in upper levels of the atmosphere (mesosphere and lower thermosphere, ~80-120 km, and stratosphere) \cite{Sakazaki} and also have an expression in the troposphere and at the surface \cite{Covey}. Indeed, for small perturbations, the linearized momentum equations for linear waves in a motionless atmosphere \cite{Holton-UL} have the form of eq. \eqref{p}, without the advective term.\\
\noindent
{\bf 6.} As shown in \cite{JC-O-Z}, eqs. (\ref{tau}-\ref{ro}) can be compared to the steady state version of shallow water equations proposed in \cite{Matsuno,Gill} for the atmospheric response to diabatic forcing confined to the tropics. This is another atmospheric phenomenon that can be described by our model.\\
\noindent
{\bf 7.} The simplified simulation above neglects the advective terms (that are not necessarily small in the Earth's atmosphere), the effect of the underlying geography and interactions with other atmospheric layers. By ignoring the vertical dimension we neglect the hydrostatic approximation, a crucial aspect for the quasi-geostrophic equations, which include the vertical advection of potential temperature required for the description of large-scale dynamics of a static stable atmosphere \cite{Phillips}. 

Including nonlinearities, geographic features and multiple layers would make the integration more challenging and the numerical approach unavoidable. These questions are left for future investigation.

{\bf Acknowledgments} We thank F. Canfora, N. Donoso, M. Rojas, R. Rondanelli and A. Sep\'ulveda, for enlightening discussions and comments. We specially thank F. Novaes for interesting critical comments, discussions and suggestions, C. Mart\'{\i}nez for his helpful advice with Mathematica, and R. Barriga for invaluable help with the programming and much more. This work has been partially supported by Fondecyt grants 1180368 and 11170486, and FONDAP 15110009.


\begin{thebibliography}{}
\bibitem{F1} If the emission is not proportional to $\tau$, like in the Stefan-Boltzmann law $E_{out}=\sigma(\tau/\rho)^4$, the formulas below would remain basically unchanged with redefined values for $\lambda$ and $\tau_0$.
\end{thebibliography}
\end{document}